\documentclass[aps,prl,onecolumn,groupedaddress]{revtex4-1}
\usepackage{graphicx}
\usepackage{here}
\usepackage{xspace}
\usepackage{amsmath}
\usepackage{amssymb}
\begin{document}
\newcommand{\sig}{$\sigma(\omega)$\xspace}
\newcommand{\Tafo}{$T_\mathrm{AFO}$\xspace}
\newcommand{\wpc}{$\omega_\mathrm{p, Coh\,}$\xspace}
\newcommand{\wpic}{$\omega_\mathrm{p, Inc\,}$\xspace}
\newcommand{\wspc}{$\omega^2_\mathrm{p, Coh\,}$\xspace}
\newcommand{\wspic}{$\omega^2_\mathrm{p, Inc\,}$\xspace}
\title{Carrier localization due to local magnetic order induced by magnetic impurities in Ba(Fe$_{1-x}$TM$_x$)$_2$As$_2$ (TM = Mn and Cr) as seen via optical spectra}

\author{T.~Kobayashi}
 \author{M.~Nakajima}
 \author{S.~Miyasaka}
 \author{S.~Tajima}
 \affiliation{Department of Physics, Osaka University, Osaka 560-0043, Japan.}

\date{\today}

\draft

\begin{abstract}
The charge dynamics of Ba(Fe$_{1-x}$TM$_x$)$_2$As$_2$ (TM = Mn and Cr) has been investigated by optical spectroscopy. 
It was found that in addition to the strong suppression of the coherent charge transport, the magnetic impurity induces a novel spectral feature related to the carrier localization in the far-infrared region above the antiferromagnetic transition temperature. 
We attribute it to the cooperative effect between conduction electrons and local magnetic order induced by magnetic impurities. 
The present results demonstrate that Mn and Cr are not conventional magnetic pair-breakers in iron pnictides.
 
\end{abstract}


\maketitle
\section{I. INTRODUCTION}
In the strongly correlated materials, the electronic fluctuation which derives from spin, orbital, charge, and lattice degrees of freedom plays a key role for the emergence of quantum phenomena, such as  ferroelectricity, colossal magnetoresistance, and high-temperature superconductivity (SC) \cite{Tokura}. 
These electronic fluctuations appear when multiple ordered phases compete with each other. 
The effects of impurities introduced into such an electronic fluctuating state are nontrivial. 
In some cases, magnetism is produced around impurity element and interacts with conduction electrons as in Zn-doped cuprates \cite{Adachi} and Cr/Al-doped manganites \cite{Kimura, Nair}.
Thus, the doped element acts as more than a simple potential scatterer in these strongly correlated electron systems. 
In other words, doped impurities possibly uncover the unusual nature of electronic background.

In iron-based superconductors, it is suggested that magnetic or orbital fluctuation plays an important role. 
SC appears when the antiferromagnetic-orthorhombic (AFO) phase in a parent compound, such as LaFeAsO and BaFe$_2$As$_2$ (Ba122), is suppressed by chemical substitution \cite{Stewart, Hirschfeld}. 
However, Ba(Fe$_{1-x}$TM$_{x}$)$_2$As$_2$ (TM=Mn, Cr; TM-Ba122) is an exceptional case that SC is not induced by chemical substitution. 
While many studies indicated that holes are not doped by Mn substitution~\cite{Kobayashi,Suzuki,Texier,Rienks}, it was confirmed that Cr substitution can introduce holes into the system~\cite{Kobayashi}. 
In contrast to the case of electron-doped Co-Ba122, where the magnetostructural transition temperature \Tafo is suppressed and the superconducting phase appears with Co substitution for Fe \cite{Ni}, in Mn/Cr-Ba122, \Tafo is suppressed by Mn/Cr substitution but a new magnetic phase appears above a critical concentration instead of the superconducting phase \cite{Thaler, Kim, Tucker, Inosov, Sefat, Marty}.
It is likely that the magnetism of Mn/Cr disturbs the appearance of superconductivity~\cite{Fernandes-Mills,Wang2014}. 
However, the effects of magnetic impurities on charge dynamics are still unclear, which is important to discuss the characteristic electronic state of this material. 

Optical spectroscopy is a useful bulk-sensitive and energy-resolved probe for investigating the electronic structure and charge dynamics \cite{Basov, Basov2, Charnukha, Tajima}.
The previous optical studies of Co-Ba122 revealed that multiple carriers contribute to low-energy charge dynamics, and an antiferromagnetic gap opens in a portion of Fermi surfaces below \Tafo, while this gap is diminished with Co substitution \cite{Hu, Nakajima, Moon}. 
In order to find the origin of different substitution effects of Co- and Mn/Cr-Ba122, it is worth to compare the low-energy charge dynamics of these systems.

In this work, we measured optical spectra of Mn/Cr-Ba122 single crystal and compared the results with those of Co-Ba122. 
We demonstrated that the coherent charge dynamics is strongly suppressed by magnetic impurity such as Mn and Cr, while it is robust against nonmagnetic impurity. 
In addition, we found an anomalous spectral feature above \Tafo in Mn/Cr-Ba122, which was not observed in Co-Ba122. 
We attribute this anomalous feature to the collective effect between conduction electrons and magnetic impurities. 

\section{II. EXPERIMENTAL METHODS}
Single crystals of Ba(Fe$_{1-x}$TM$_x$)$_2$As$_2$ ($x=0.04, 0.08$, and $0.12$ for TM=Mn, and $x=0.02, 0.06, 0.09$, and $0.13$ for Cr) were grown by a self flux method as described in Ref.~\onlinecite{Kobayashi}. 
The compositions of the grown crystals were determined by scanning electron microscopy-energy dispersion x-ray (SEM-EDX) analysis. 
Electrical resistivity was measured by a standard four-probe method.

In-plane optical reflectivity was measured in an energy range of 50--25000~cm$^{-1}$ at various temperatures from 5 to 300~K using a Fourier-transform-type spectrometer. 
The samples were polished using alumina powder to obtain flat and shiny surfaces. 
Although it was pointed out that optical spectra suffer from polishing effect~\cite{Moon2013}, in some cases, the polished surface gives the same results as those on a cleaved surface~\cite{Nakajima,Uykur}. 
Moreover, in the present study, we mainly focus on the impurity concentration dependence of the spectra, rather than the absolute values of conductivity. 
When discussing the systematic change of the spectra with impurity substitution, the polishing effect on it would be regarded as minor. 

The reflectivity spectrum for the ultraviolet and vacuum-ultraviolet region was measured at room temperature at BL7B of UVSOR, Institute for Molecular Science at Okazaki, Japan.
The optical conductivity \sig was obtained from the reflectivity spectra using the Kramers-Kronig transformation.
We used the Hagen-Rubens formula for the low-energy extrapolation and the free electron response ($\omega^{-4}$) for the higher energy extrapolation.

\section{III. RESULTS AND DISCUSSION}

\begin{figure}
\begin{center}
\includegraphics[width=0.45\textwidth]{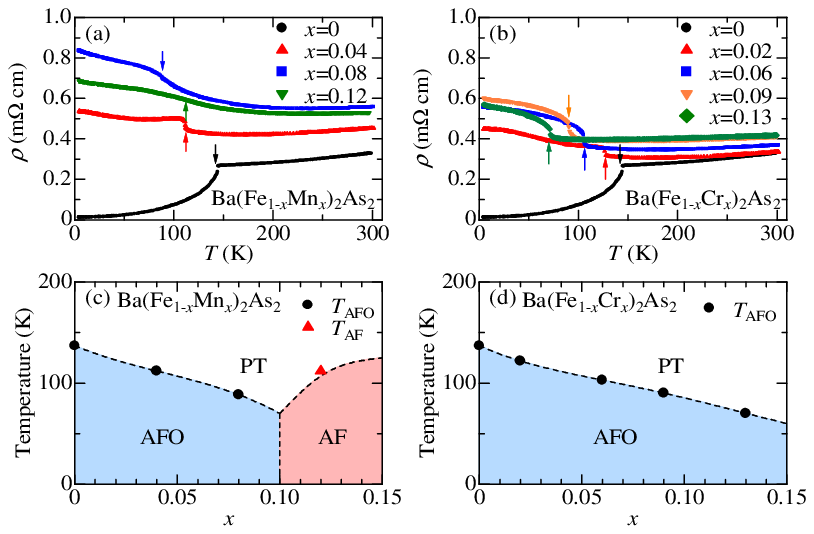}\\
\end{center}
\caption{(Color online) $T$ dependence of $\rho(T)$ for (a) Ba(Fe$_{1-x}$Mn$_x$)$_2$As$_2$ and (b) Ba(Fe$_{1-x}$Cr$_x$)$_2$As$_2$. Arrows indicate $T_\mathrm{AFO}$ and $T_\mathrm{AF}$. The electronic phase diagram of (c) Ba(Fe$_{1-x}$Mn$_x$)$_2$As$_2$ and (d) Ba(Fe$_{1-x}$Cr$_x$)$_2$As$_2$ are determined from the $T$ derivative of $\rho(T)$. PT, AFO, and AF represent the paramagnetic-tetragonal, antiferromagnetic-orthorhombic, and antiferromagnetic-tetragonal phase, respectively. We identified AFO and AF phase, based on the previous works~\cite{Kim,Marty}.}
\label{fig1}
\end{figure}
All the detailed reflectivity and conductivity spectra are shown in the Appendix, including the precise temperature dependence as well as the reflectivity data in a wide frequency range. 
Here we focus on the far-infrared conductivity spectra where the impurity effect is remarkable. 

Figures \ref{fig1}(a) and \ref{fig1}(b) show the temperature ($T$) dependence of in-plane resistivity $\rho(T)$ of Mn- and Cr-Ba122, respectively. 
For BaFe$_2$As$_2$, $\rho(T)$ shows a metallic $T$ dependence with an abrupt drop at $T_\mathrm{AFO}=138~\mathrm{K}$. 
With Mn/Cr substitution, the $\rho(T)$ value at room $T$ increases, which is different from the case of Co substitution where the $\rho(T)$ value decreases with doping level \cite{Nakajima}. 
In BaFe$_2$As$_2$, $\rho(T)$ initially decreases and then shows an abrupt drop below \Tafo, while in Mn/Cr-Ba122 $\rho(T)$ shows a jump at \Tafo and then increases with lowing $T$. 
As Mn/Cr content increases, the jump feature is smeared out. 
Figures \ref{fig1}(c) and \ref{fig1}(d) show the electronic phase diagrams of Mn- and Cr-Ba122, respectively. 
$T_\mathrm{AFO}$ and $T_\mathrm{AF}$ were determined from $\rho(T)$, which are consistent with the values in the previous studies \cite{Kim, Marty}. 
Mn/Cr substitution suppresses the AFO phase, but the SC does not appear in contrast to the case of Co-Ba122 where SC and AFO order are coexisting when \Tafo becomes lower than 100 K~\cite{Nakajima,Ishida}. 
Above $x=0.10$ for Mn-Ba122, the antiferromagnetic-tetragonal phase appears instead of the AFO one \cite{Kim}.

\begin{figure}
\begin{center}
\includegraphics[width=0.45\textwidth]{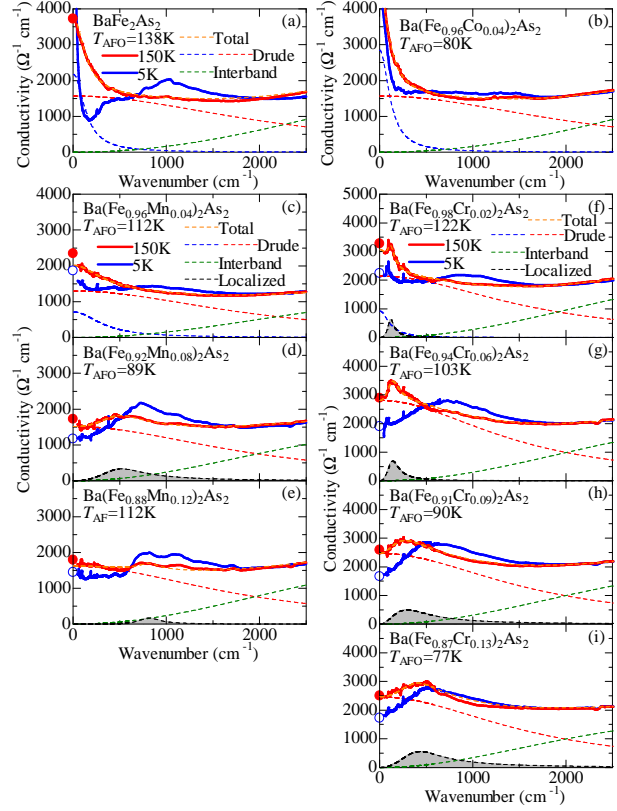}\\
\end{center}
\caption{(Color online) The \sig for (a)--(b) Ba(Fe$_{1-x}$Co$_x$)$_2$As$_2$, (c)--(e) Ba(Fe$_{1-x}$Mn$_x$)$_2$As$_2$ ($x = 0.04, 0.08$, and $0.12$), and (f)--(i) Ba(Fe$_{1-x}$Cr$_x$)$_2$As$_2$ ($x = 0.02, 0.06, 0.09$, and $0.13$) at 150~K (red) and 5~K (blue). 
The $dc$ conductivity at 150 K and 5 K estimated from the resistivity data are plotted as closed and open circles on the vertical axes, respectively. 
The dashed lines show the fitting results for the spectra at 150~K with the Drude-Lorentz formula. 
The unusual peaks observed for Mn/Cr-Ba122 are shaded for the clarity. 
The results for Co-Ba122 ($x=0$ and $0.04$) in the previous study \cite{Nakajima} are replotted in panels (a) and (b).}
\label{fig2}
\end{figure}
The \sig spectra of Co-Ba122 and Mn/Cr-Ba122 at 150~K ($>$\Tafo) and 5~K ($<$\Tafo) are shown in Fig.~\ref{fig2}. 
The $dc$ conductivity are also indicated in the figure. 
For $x=0$ [Fig.~\ref{fig2}(a)], \sig at 150~K shows a peak at $\omega=0$ and a long tail extending to $\sim2000$~cm$^{-1}$. 
Below \Tafo, \sig is suppressed below 650~cm$^{-1}$ and a broad peak appears at 1000~cm$^{-1}$, indicating the formation of the AFO gap, while the residual Drude peak gets sharper.
For 4\% Co-Ba122 [Fig.~\ref{fig2}(b)], the overall feature of \sig both above and below \Tafo is similar to that of BaFe$_2$As$_2$ except for the reduced AFO gap energy.
By contrast, with Mn/Cr substitution, the $\omega=0$ peak is strongly suppressed both above and below \Tafo, as shown in Figs.~\ref{fig2}(c)--\ref{fig2}(i). 
Instead, an additional peak structure appears at 300--800~cm$^{-1}$ at 150~K. 
Interestingly, the peak position shifts to higher energy with increasing Mn/Cr concentration.  
At 5~K, the AFO gap is formed at 500--1000~cm$^{-1}$ and the gap energy is monotonically reduced with Mn/Cr substitution, which is similar to the case of Co-substitution.

In the present work, we focus on the charge dynamics in the paramagnetic phase and its change with Mn and Cr substitution. 
To this end, we fit \sig at $T=150$~K ($>$\Tafo) with the Drude-Lorentz model. 
\begin{displaymath}
\sigma(\omega)=\sum^{}_\mathrm{i}\frac{\omega^2_\mathrm{p,i}}{4\pi}\frac{\tau_\mathrm{i}}{(\omega\tau_\mathrm{i})^2+1} + \sum^{}_\mathrm{j}\frac{S_\mathrm{j}^2}{4\pi}\frac{\tau_\mathrm{j}\omega^2}{\tau_\mathrm{j}^2(\omega_\mathrm{j}^2-\omega^2)^2+\omega^2} 
\end{displaymath}
The first term describes the response from the free carriers and the second term describes the interband transitions. 
Here, $\omega_\mathrm{p,i}$ is the plasma frequency and $1/\tau_\mathrm{i}$ is the scattering rate of the carriers. 
$\omega_\mathrm{j}$, $1/\tau_\mathrm{j}$, and $S_\mathrm{j}$ are the resonance frequency, line width, and strength of each Lorentz oscillator, respectively.
As shown in Fig.~\ref{fig2}(a), the low-energy spectrum at $x=0$ can be fitted by using two (narrow and broad) Drude terms and one interband transition term with a peak at around 5000~cm$^{-1}$. 
We call the narrow Drude term the coherent term and the broad one the incoherent term, following the previous study \cite{Nakajima}. 
For $x=0.04$ in Mn-Ba122 and $x=0.02$ in Cr-Ba122 [Figs.~\ref{fig2}(c) and \ref{fig2}(f), respectively], two Drude terms are required to reproduce the observed \sig,  but the weight of the coherent term is reduced and its width is broadened. 
With further Mn/Cr substitution [Figs.~\ref{fig2}(d), \ref{fig2}(e) and \ref{fig2}(g)--\ref{fig2}(i)], a clear $\omega=0$ peak is absent, which makes the two-Drude analysis difficult, and thus \sig has been fitted by one incoherent Drude term.
\begin{figure}
\begin{center}
\includegraphics[width=0.45\textwidth]{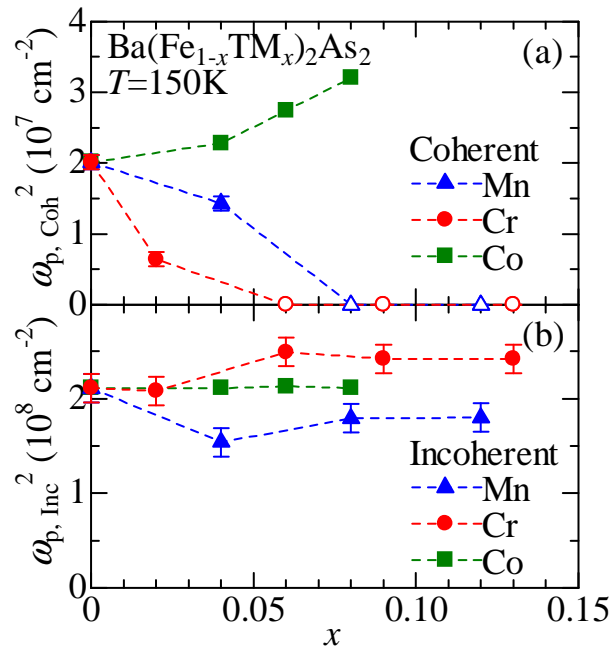}\\
\end{center}
\caption{(Color online) The doping dependence of the spectral weight of Ba(Fe$_{1-x}$TM$_x$)$_2$As$_2$ [TM=Mn (blue triangles), Cr (red circles), and Co (green squares)] for (a) coherent part ($\omega_\mathrm{p, Coh}^2$) and (b) incoherent part ($\omega_\mathrm{p, Inc}^2$). Here, $\omega_\mathrm{p, Coh}$ and $\omega_\mathrm{p, Inc}$ are plasma frequency of coherent and incoherent Drude terms. 
$\omega_\mathrm{p, Coh}$ above $x=0.06$ for Cr-Ba122 and $0.08$ for Mn-Ba122 (open symbols) were assumed to be zero because it is difficult to distinguish the coherent and the incoherent term. 
The results for TM=Co in the previous study \cite{Nakajima} are also replotted.}
\label{fig3}
\end{figure}

 In Fig. \ref{fig3}, we plot the doping dependence of the spectral weight of the coherent and the incoherent Drude terms (\wspc and \wspic, respectively) at 150~K estimated by the fitting in Fig.~\ref{fig2}. 
The uncertainty of Drude-Lorentz analysis is indicated by an error bar in Fig. 3. 
 As it is clearly seen in Fig.~\ref{fig3}(a), \wspc is strongly suppressed by Mn/Cr substitution and vanishes above $x=0.08\,(0.06)$ for Mn (Cr).
 On the other hand, the spectral weight of the incoherent term (\wspic) slightly increases with Cr substitution, while it decreases with Mn substitution [Fig.~\ref{fig3}(b)].
 The doping dependence of spectral weight of Mn/Cr-Ba122 is different from that of Co-Ba122 where \wspc monotonically increases with Co substitution and \wspic is almost doping independent.
 Since \wspic is one order larger than \wspc, it turns out that the total Drude spectral weight is governed by the incoherent term in Mn/Cr-Ba122.
If we assume that the effective mass does not change with impurity substitution, the present result indicates that the total carrier number increases with Cr substitution, while it decreases with Mn substitution. 
This is consistent with the previous observation in Hall effect and photoemission experiments that the hole carriers are doped in Cr-Ba122 but not in Mn-Ba122~\cite{Kobayashi, Suzuki}. 
In spite of the increase of carrier number,  Mn/Cr substitution strongly suppresses the coherent carrier transport, and consequently the incoherent Drude component dominates the $dc$ conductivity in Mn/Cr-Ba122. 
This is consistent with the $\rho(T)$ behavior of Mn/Cr-Ba122 in Fig. 1, which indicates that the charge transport gets incoherent with substitution. 
It implies that the doped Mn/Cr acts as a stronger scatterer than Co, and the carrier transport loses its coherence by Mn/Cr substitution. 
\begin{figure}
\begin{center}
\includegraphics[width=0.45\textwidth]{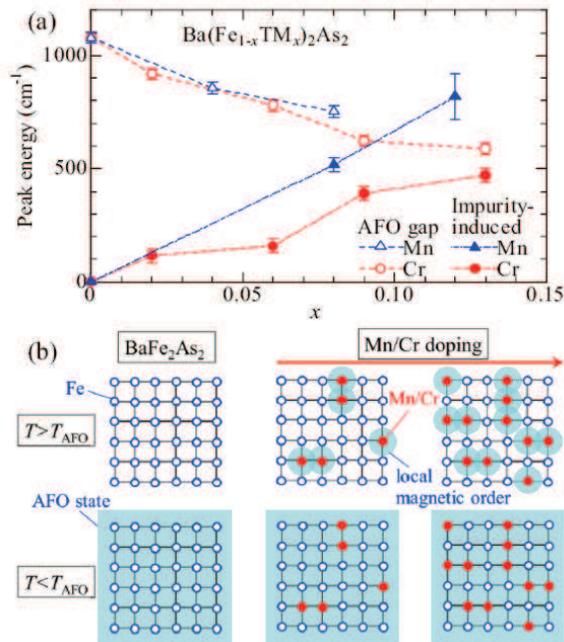}\\
\end{center}
\caption{(Color online) (a) The doping dependence of the AFO-gap (open) and the localized-peak (closed) energy 
for Ba(Fe$_{1-x}$TM$_x$)$_2$As$_2$ [TM = Mn (blue triangles) and Cr (red circles)]. 
The gap energy is estimated by the fitting with the Lorentz term. 
The gap energy at 5~K for $x=0.12$ of Mn-Ba122 is not included because the different magnetic phase is realized \cite{Kim}.
(b) Schematic illustration of Fe (blue open circles) plane in pure and Mn/Cr-Ba122. 
Above \Tafo (upper panels), the doped Mn/Cr (red closed circles) induces the local magnetic ordered regions (light blue circles) around itself, and their area increases with the doping level.
Below \Tafo (lower panels), the long range order (light blue area) dominates the system. 
As a result, the signature of local magnetic order becomes unclear in the optical spectra.}
\label{fig4}
\end{figure}

Another and more remarkable feature induced by Mn/Cr substitution is a far-infrared peak structure that appears in the low-energy \sig in the paramagnetic state. 
Figure \ref{fig4}(a) shows the doping dependence of the new peak energy at 150~K and the AFO gap peak energy at 5~K. 
The peak structure at a finite frequency is a signature of carrier localization. 
One of the possible origins is a precursor of the AFO gap. 
However, as seen in Fig.~\ref{fig4}(a), the new far-infrared peak shifts to higher energy with increasing Mn/Cr concentration, which is opposite to the behavior of the AFO gap below \Tafo. 
Therefore, we conclude that these two peaks are different in origin. 
The second possibility is the charge excitation from the doped Mn and Cr $3d$ states. 
However, it can be ruled out because the peak energy changes with $x$. 
Moreover, the resonance photoemission measurements revealed that the partial density of state of doped Mn and Cr exists at 1--10~eV below the Fermi energy \cite{Suzuki}, which is far from the energy expected from the observed peak ($\sim50$~meV).

Another possible origin is the disorder effect.
Actually, a similar low-energy peak is sometimes observed in disordered systems, such as Zn-doped, Ni-doped, and He-irradiated cuprate superconductors \cite{Basov3, Homes, Basov4}. 
Thus the present results seem to be attributed to the disorder effect by Mn/Cr-substitution. 
However, this cannot explain why such a disorder effect is not observed in Co-Ba122 where disorders are also introduced into Fe planes \cite{Nakajima}.
The difference of Co and Mn/Cr is that the former is nonmagnetic and the latter is magnetic impurity \cite{Kim, Marty}. 
Therefore, the interplay between the conduction electrons and the magnetic impurities should be considered.
A representative interaction between the magnetic impurity and the conduction electron is that for the Kondo effect. 
Kondo materials or heavy-fermion systems, such as YbB$_{12}$ and CeCu$_2$Si$_2$ \cite{Okamura, Sichelschmidt}, show a peak structure of \sig in the far-infrared energy region.  
The $T$ dependence of $\rho(T)$ of Mn/Cr-Ba122 shows a small upturn above \Tafo, which may be explained by the Kondo effect \cite{Urata}. 
However, the dynamical mean-field theory \cite{Georges} suggests that a multiband system with large Hund's coupling $J$ has a significantly reduced Kondo $T$, and a large $J$ value is required to explain the observed electronic structure of Mn-Ba122 \cite{Suzuki}.
This is inconsistent with the present observation that the peak structure in \sig is already present above \Tafo which is much higher than the expected Kondo $T$.

The compelling origin is the cooperative effect between magnetic impurities and conduction electrons. 
Gastiasoro and Andersen have theoretically proposed that the magnetic moment of doped impurities interacts with the magnetic fluctuation of itinerant electrons and induces a local magnetic order around the impurities even in the paramagnetic state of iron pnictides \cite{Gastiasoro}. 
A schematic illustration is shown in Fig.~\ref{fig4}(b). 
A small amount of magnetic impurities induces a local N\'eel-type magnetic order above \Tafo. 
At the higher concentrations of magnetic impurity, the stripe-type magnetic order is dominant even above \Tafo due to the cooperative effect of the magnetic impurities and conduction electrons. 
This theory can explain the previous experimental observation in Mn-Ba122 that N\'eel- and stripe-type magnetic orders coexist and a tetragonal-stripe type magnetic order emerges in the higher doping range \cite{Kim, Tucker}.

Such a cooperative effect can cause the strong suppression of the coherent carrier transport, and it is one of the possible origins for the localization peak observed above \Tafo, although Gastiasoro and Andersen did not calculate any optical response in their work \cite{Gastiasoro}. 
The increase of the peak energy with $x$ reflects the characteristic mechanism of carrier localization in which interaction strength between local magnetic moments around impurities increases with impurity concentration. 
In this scenario, the fact that the increasing rate of the impurity-induced peak energy is larger in Mn-Ba122 than in Cr-Ba122 (Fig.~\ref{fig4}(a)) implies the difference of interaction strength between these systems. 
This difference results in the different phase diagram (Fig. 1) where antiferromagnetic-tetragonal order appears above $x=0.10$ in Mn-Ba122 \cite{Kim}, while the original antiferromagnetic-orthorhombic order persists up to $x=0.30$ in Cr-Ba122 \cite{Marty}. 
However, so far there has been no theoretical calculation of the charge dynamics as a function of impurity concentration in iron pnictides. 
More detailed theoretical study is required. 

The localization peak seen above \Tafo becomes invisible below \Tafo \cite{comment}, and instead the AFO gap structure appears as shown in Figs.~\ref{fig2}(d)--\ref{fig2}(i).
This is because the long-range antiferromagnetic order dominates the system as illustrated in Fig.~\ref{fig4}(b). 
In the AFO phase, the coherent Drude term of Mn/Cr-Ba122 is more strongly suppressed than that of Co-Ba122, which demonstrates that Mn and Cr act as strong scattering centers also in the AFO state. 

The loss of carrier coherence due to magnetic impurities (Mn and Cr) prevents the appearance of SC, although the conventional magnetic pair breaking could be another source for non-SC \cite{Fernandes-Mills, Wang2014}. 
Then, it is understandable that SC is not observed in Mn/Cr-Ba122 even though \Tafo is well suppressed by Mn/Cr substitution, in contrast to the case of Co-Ba122 \cite{Ni, Nakajima, Nakajima2}, where the coherent Drude component well survives with Co substitution. 

\section{IV. CONCLUSION}
We performed \sig measurements on Mn/Cr-Ba122. 
The coherent component of \sig is strongly suppressed by magnetic Mn/Cr-substitution both in the paramagnetic and the AFO states.
This result demonstrates that the coherent carriers are strongly scattered by the magnetic impurity (Mn/Cr) although carrier scattering by the nonmagnetic impurity (Co) is very weak in iron pnictides. 
The rapid suppression of coherent carries transport is one of the factors to hinder the appearance of SC. 
In the paramagnetic phase above \Tafo, we found the magnetic impurity-induced peak in the far-infrared spectra that shifts to higher energy with increasing Mn/Cr concentration. 
We attribute this anomalous carrier localization to the cooperative effect between conduction electrons and magnetic impurities.

\
\section{ACKNOWLEDGMENTS}
This work was supported by Grants-in-Aid for Scientific Research from MEXT and JSPS, Japan. T.~K.\, acknowledges the Grant-in-Aid for JSPS Fellows.


\section{V. APPENDIX}
Here we provide all the detailed results of optical reflectivity and conductivity to supplement the main text. 
Figures 5 and 6 show the precise temperature dependence of reflectivity and conductivity spectra for Mn- and Cr-Ba122, respectively. 
Figure 7 shows the reflectivity data for Mn- and Cr-Ba122 in a wide frequency range. 

\begin{figure}[h]
\begin{center}
\includegraphics[width=0.45\textwidth]{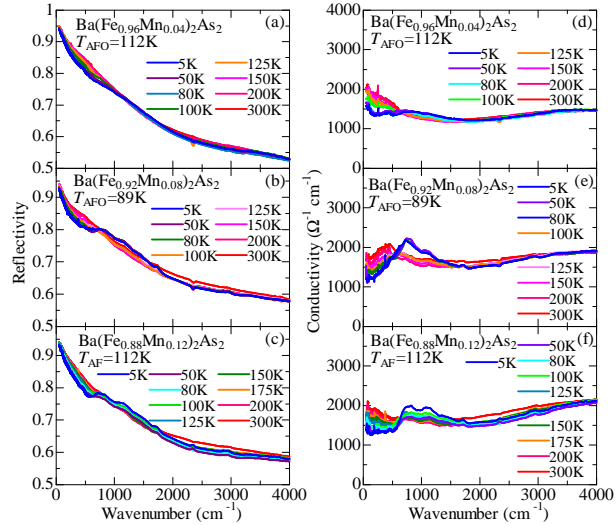}\\
\end{center}
\caption{(Color online) Optical reflectivity ((a)-(c)) and conductivity ((d)-(f)) of Ba(Fe$_{1-x}$Mn$_x$)$_2$As$_2$ ($x = 0.04, 0.08$, and $0.12$) at various temperatures.}
\label{fig5}
\end{figure}
\begin{figure}[h]
\begin{center}
\includegraphics[width=0.45\textwidth]{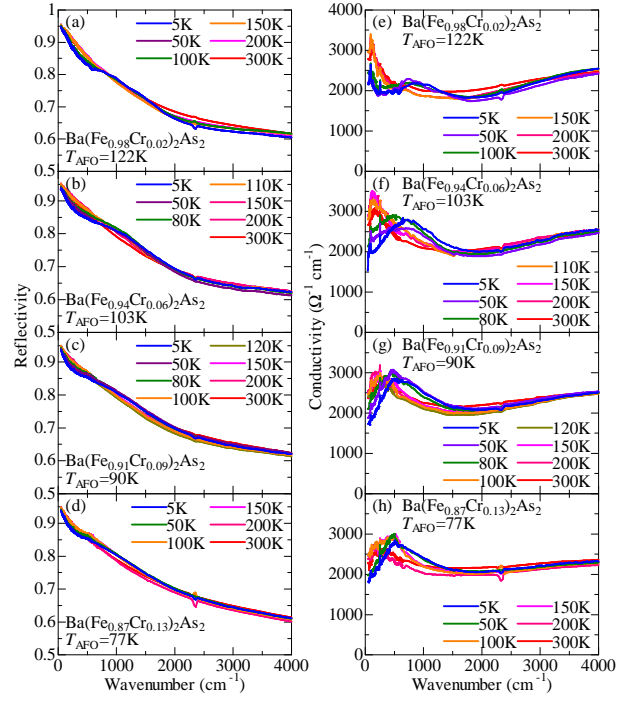}\\
\end{center}
\caption{(Color online) Optical reflectivity ((a)-(d)) and conductivity ((e)-(h)) of Ba(Fe$_{1-x}$Cr$_x$)$_2$As$_2$ ($x = 0.02, 0.06, 0.09$, and $0.13$) at various temperatures.}
\label{fig6}
\end{figure}
\
\newpage
\
\begin{figure}[h]
\begin{center}
\includegraphics[width=0.45\textwidth]{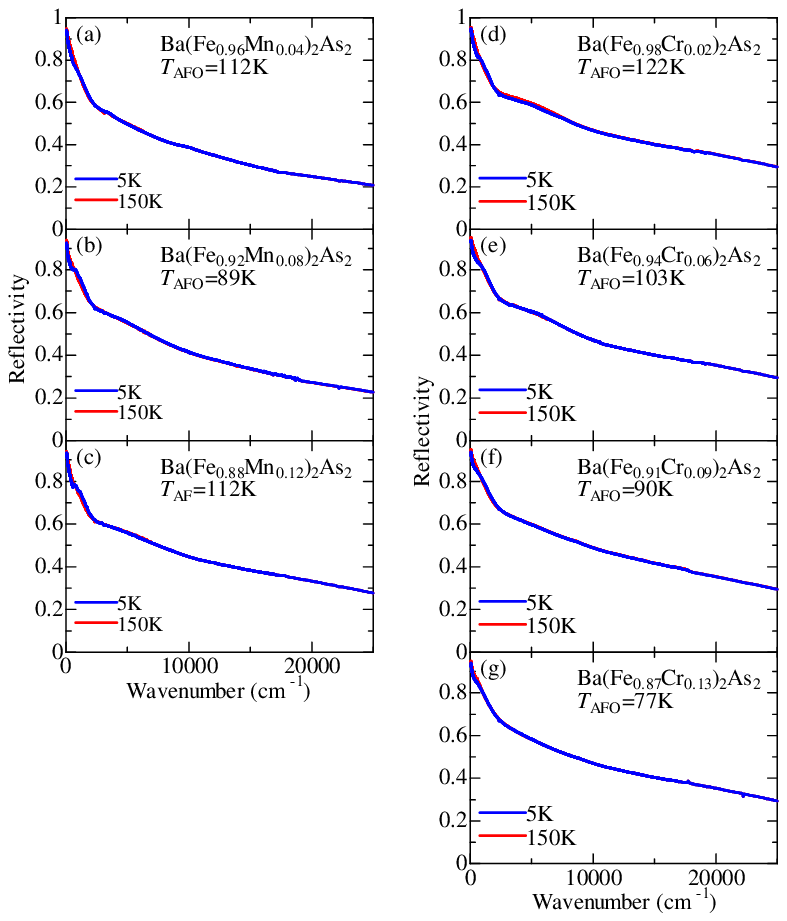}\\
\end{center}
\caption{(Color online) Optical reflectivity in wide frequency range (up to 25000 cm$^{-1}$) of Ba(Fe$_{1-x}$Mn$_x$)$_2$As$_2$ ($x = 0.04, 0.08$, and $0.12$) and Ba(Fe$_{1-x}$Cr$_x$)$_2$As$_2$ ($x = 0.02, 0.06, 0.09$, and $0.13$) at 5 K and 150 K.}
\label{fig6}
\end{figure}

\newpage



\begin{thebibliography}{99}
\bibitem{comment}One may see the low $\omega$ peak also below \Tafo in Cr-Ba122 with $x=$0.02 and 0.06. However, we cannot assign it to the localization peak. First, we have to remind that there is a phonon peak ascribed to the Ba vibrational mode around 90 cm$^{-1}$ which survives even if the localization peak at 100 cm$^{-1}$ disappears below \Tafo. Second, since the spectra below 80-90 cm$^{-1}$ is close to the limit of our measurement range, sometimes the oscillating noise is overlapped with the spectrum. This could create an apparent peak. Third, assuming that the lowest $\omega$ peak below 80 cm$^{-1}$ is the localization peak, we find that it shifts with temperature from 100 cm$^{-1}$ to $\sim$80 cm$^{-1}$. This is unlikely. From all these reasons, we think that there is no clear signature of localization peak below \Tafo in the optical spectra, although the local magnetic order may remain even below \Tafo.
\end{thebibliography}
\end{document}